\pdfoutput=1

\documentclass[sigconf,natbib=true,dvipsnames,svgnames]{acmart}

\renewcommand\footnotetextcopyrightpermission[1]{} 
\usepackage[T1]{fontenc}

\usepackage[hang,flushmargin]{footmisc}

\usepackage[T1]{fontenc}    
\usepackage{xurl}            
\usepackage{booktabs}       
\usepackage{amsfonts}       
\usepackage{nicefrac}       
\usepackage{microtype}      
\usepackage{xcolor}         
\usepackage{amsmath}
\usepackage{enumitem}
\usepackage{graphicx}       
\usepackage{caption}
\usepackage{subcaption}
\usepackage{threeparttablex} 
\usepackage[export]{adjustbox} 
\usepackage{multirow}

\makeatletter
\@namedef{ver@everyshi.sty}{}
\makeatother

\newenvironment{customlegend}[1][]{%
    \begingroup
    \csname pgfplots@init@cleared@structures\endcsname
    \pgfplotsset{#1}%
}{%
    \csname pgfplots@createlegend\endcsname
    \endgroup
}%
\def\addlegendimage{\csname pgfplots@addlegendimage\endcsname}

\usepackage{pgfplots, pgfplotstable}
\pgfplotsset{compat=newest}
\usepgfplotslibrary{fillbetween}

\newcommand{\vitbc}{Blue}
\newcommand{\vitlc}{PineGreen}
\newcommand{\vithc}{Salmon}
\newcommand{\vitgc}{RubineRed}
\newcommand{\bmvc}{Black}

\newcommand{\openclipc}{Blue}
\newcommand{\clipc}{Salmon}
\newcommand{\flavac}{PineGreen}
\newcommand{\blipc}{RubineRed}

\pgfmathsetmacro{\stdgrad}{30}

\tikzset{every mark/.append style={solid}}
\pgfplotsset{
	grid=both, width=\linewidth, try min ticks=5,
	legend cell align=left, legend style={fill opacity=0.8},
	ylabel near ticks,
    xlabel near ticks,
    every tick label/.append style={font=\footnotesize},
}

\pgfplotsset{
    vitbplot/.style={thick, color=\vitbc, mark=triangle*, mark size=2pt},
    vitlplot/.style={thick, color=\vitlc, mark=oplus*, mark size=2pt},
    vithplot/.style={thick, color=\vithc, mark=pentagon, mark size=2pt},
    vitgplot/.style={thick, color=\vitgc, mark=x, mark size=2pt},
    bmvTplot/.style={thick, color=\bmvc, mark=o, mark size=2pt},
    openclipplot/.style={color=\openclipc, mark=triangle*, mark size=2pt, only marks},
    openclipFplot/.style={color=\openclipc, mark=o, mark size=2pt, only marks},
    clipplot/.style={only marks, color=\clipc, mark=pentagon, mark size=2pt},
    flavaplot/.style={only marks, color=\flavac, mark=x, mark size=2pt},
    blipplot/.style={only marks, color=\blipc, mark=square, mark size=2pt},
    equation/.style={thick, color=Black, dotted, domain=0:100},
    equationD/.style={thick, color=Grey, dashed, domain=0:100},
}

\usepackage{colortbl}
\definecolor{Gray}{gray}{0.5}
\definecolor{GrayBG}{gray}{0.95}

\newcommand\ignore[1]{}

\newcommand\blue[1]{\textcolor{blue}{#1}}

\let\oldFootnote\footnote
\newcommand\nextToken\relax

\renewcommand\footnote[1]{%
    \oldFootnote{#1}\futurelet\nextToken\isFootnote}

\newcommand\isFootnote{%
    \ifx\footnote\nextToken\textsuperscript{,}\fi}

\title{AToMiC: An Image/Text Retrieval Test Collection\\ to Support Multimedia Content Creation}

\author{Jheng-Hong Yang}
\affiliation{University of Waterloo \country{Canada}}

\author{Carlos Lassance}
\affiliation{Naver Labs Europe \country{France}}

\author{Rafael S. Rezende}
\affiliation{Naver Labs Europe \country{France}}

\author{Krishna Srinivasan}
\affiliation{Google Research \country{United States}}

\author{Miriam Redi}
\affiliation{Wikimedia Foundation \country{United Kingdom}}

\author{Stéphane Clinchant}
\affiliation{Naver Labs Europe \country{France}}

\author{Jimmy Lin}
\affiliation{University of Waterloo \country{Canada}}

\begin{document}

\begin{abstract}
This paper presents the AToMiC (\textbf{A}uthoring \textbf{To}ols for \textbf{M}ult\textbf{i}media \textbf{C}ontent) dataset, designed to advance research in image/text cross-modal retrieval. 
While vision--language pretrained transformers have led to significant improvements in retrieval effectiveness, existing research has relied on image--caption datasets that feature only simplistic image--text relationships and underspecified user models of retrieval tasks. 
To address the gap between these oversimplified settings and real-world applications for multimedia content creation, we introduce a new approach for building retrieval test collections.
We leverage hierarchical structures and diverse domains of texts, styles, and types of images, as well as large-scale image--document associations embedded in Wikipedia.
We formulate two tasks based on a realistic user model and validate our dataset through retrieval experiments using baseline models.
AToMiC offers a testbed for scalable, diverse, and reproducible multimedia retrieval research.
Finally, our dataset provides the basis for a dedicated track at the 2023 Text Retrieval Conference (TREC), and is publicly available at~\blue{\url{https://github.com/TREC-AToMiC/AToMiC}}.
\end{abstract}

\maketitle

\section{Introduction}\label{sec:intro}

Multimedia content creation requires an understanding of the connections between elements encoded in different modalities.
Frequently, visual elements such as photos, graphics, and diagrams are used to supplement textual information, to fulfill different purposes such as decorating, complementing, or transforming the meaning of the content~\cite{levin1978pictures,marsh2003taxonomy}.
However, recent research concerning multimedia information retrieval usually relies on datasets that assume a simple relationship between images and texts.

Since the great success of Contrastive Language–Image Pre-training (CLIP)~\cite{radford2021learning}, most recent research on vision--language pretrained transformers utilize image captioning datasets for retrieval benchmarking~\cite{li2021align,zhai2022lit,li2022blip,singh2022flava,yao2022filip}, e.g., COCO~\cite{lin2014microsoft} or Flickr\-30k~\cite{young2014image}.
Typically, text segments in these datasets describe the image content in a generic and literal way, like the caption ``a man on a bicycle riding next to a train'' taken from COCO, while multimedia content creation requires a richer and more precise understanding of the context, e.g., ``Larry at the door of 10 Downing Street, his official residence and workplace'', where Larry is actually referring to a cat.\footnote{\url{https://en.wikipedia.org/wiki/Larry_(cat)}}
To advance the state of the art, research communities require datasets that capture the different ways in which images and texts are connected in natural settings.
We introduce the \textbf{A}uthoring \textbf{To}ols for \textbf{M}ult\textbf{i}media \textbf{C}ontent (AToMiC) dataset to study this gap between images and texts in the context of Wikipedia.

Connections between images and texts are complex and have different purposes.
Recent work provides various kinds of image--text collections in order to study nuanced author intents of when and why images and texts are associated.
For instance, the work of~\citet{vempala-preotiuc-pietro-2019-categorizing} identifies four classes of information overlap between texts and images in Twitter posts, while~\citet{kruk-etal-2019-integrating} reveal that the intent behind Instagram posts can serve more sophisticated purposes such as irony. 
\citet{Lommatzsch2022NewsImagesAT} highlight the depiction gap between the images and their associated news article, where the images are often only loosely connected to the events described in the article.
The growing body of work on image--text relations underscores the need for more resources to advance our understanding in this area.

\begin{figure}[t]
\centering
\includegraphics[width=\columnwidth]{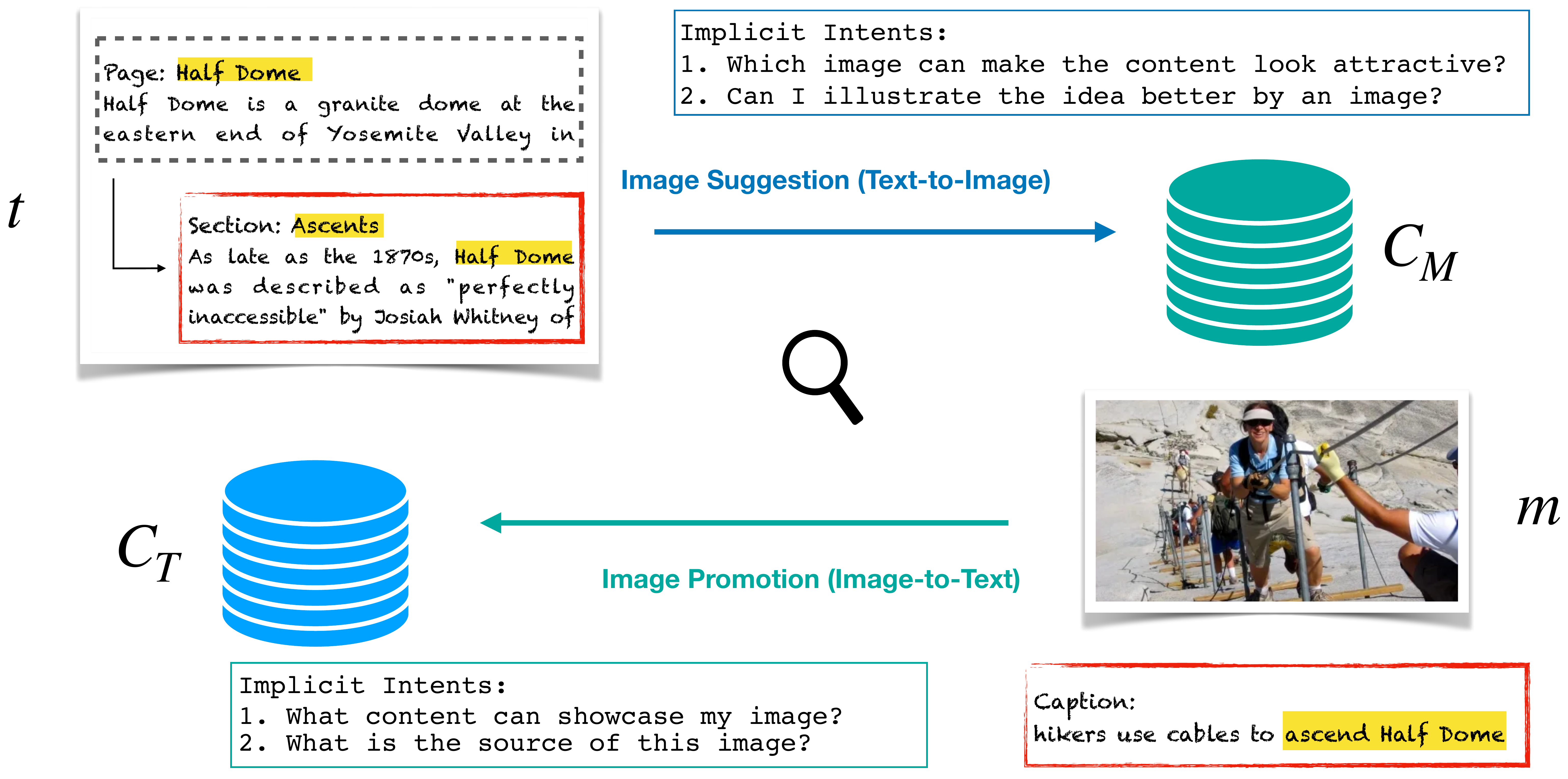}
\caption{
Our two retrieval tasks:\ image suggestion and image promotion. 
We use existing image--text associations in Wikipedia as proxy relevance labels.
}
\vspace{-0.25cm}
\label{fig:teaser}
\end{figure}

The AToMiC dataset is designed to address a real user need:\ assisting authors in multimedia content creation to enhance the appeal of web pages that are primarily textual in nature, e.g., a travel blog post~\cite{tbas_2009} or Wikipedia pages.\footnote{Short link to a related Wikimedia initiative:~\url{https://tinyurl.com/3dasup3r}}
It is a common practice for authors to make editorial decisions in selecting appropriate images (or other multimedia content such as videos) to complement the texts~\cite{xu2014lights,fujii2016enriching,bibi2020design,rama2022large}.
Moreover, the information quality of multimedia content is often related to the attached images~\cite{stvilia2005information}.
However, this editorial process is laborious and time-consuming.
Likewise, the author may not have a clear idea of what an image should represent (i.e., ignorant of relevant keywords, concepts, or related topics), and may need a system to interpret the text and suggest good image candidates.
Building assistance tools that enable efficient access to multimedia repositories could facilitate more effective content creation and iteration.

To address the aforementioned needs for a diverse and realistic test collection for image/text retrieval tasks, we introduce the AToMiC dataset, which is based on Wikipedia's text and image collections as well as their associations curated by Wikipedia editors.
The AToMiC dataset offers valuable evaluation resources, baselines, and a community for studying natural occurrences of texts and images in multimedia encyclopedic documents.
We present experiments that analyze and demonstrate the potentials of the AToMiC dataset in real-world scenarios, while also addressing its limitations such as label and popularity biases.
Specifically, we formulate two retrieval tasks and conduct a preliminary evaluation using publicly available transformer-based models.
The AToMiC dataset is publicly available and will be featured in a dedicated track at the 2023 Text Retrieval Conference (TREC). 

In summary, our contributions are listed as follows:
 \begin{enumerate}[leftmargin=*]
     \item We formulate two multimedia retrieval tasks based on real user needs:\ image suggestion and image promotion.
     \item We provide an easy-to-use dataset to support these tasks.
     The dataset is freely accessible and will be used in a TREC 2023 evaluation, together with a test set with richer annotations.
     \item We perform initial experiments and analyses to compare with the widely used COCO dataset and provide baselines to show the capabilities of AToMiC for the proposed tasks.
 \end{enumerate}

\section{Related Work}

We begin with an overview of other image/text evaluation resources and explain where AToMiC fits in this broader context:

\paragraph{ImageCLEF: 2012 $\sim$ today.}
The ImageCLEF Challenges have organized multimodal retrieval tasks on Wikipedia for several years \cite{imageclef2010_book,imageclef2010_photoretrieval,clef2011}, the latest one in 2011.
Since then, the research scope of ImageCLEF has shifted to domain-specific applications such as medical, social media, and nature applications~\cite{10.1007/978-3-031-13643-6_31}.
Our dataset also uses Wikipedia but focuses on developing a general benchmark for the systems that can process data from a wide variety of domains.

\paragraph{INEX and ImageCLEF before 2012.}
Our proposed evaluation is related to the INEX \textit{ad hoc} tracks~\cite{fuhr2007overview} and the Wikipedia challenges hosted by ImageCLEF~\cite{popescu2010overview}.
These evaluations use Wikipedia as their resource to construct test collections and formulate their tasks to adhere closely to ``classic'' TREC \textit{ad hoc} retrieval tasks.
While the INEX tracks are dedicated to developing XML retrieval systems that can process structured documents, the Wikipedia challenges focus on searching images with queries composed of various modalities.
AToMiC delivers an up-to-date and larger test collection that combines the ideas of these early initiatives, i.e., leveraging the document structures in Wikipedia and developing systems that can process various modalities.

\paragraph{FIRE: Automatic Story Illustration.}
The objective of the Automatic Story Illustration (ASI) track hosted by the FIRE workshop is to build a test collection for the task of document expansion with images, i.e., illustrate a story by retrieving images that are relevant to the text segments~\cite{ganguly2015overview}.
The ASI track also uses Wikipedia as its image and text collection.
However, it only focuses on a few topics about children’s fairy tales.
Furthermore, this test collection is not publicly available anymore.

\paragraph{Image--caption datasets.}
Recent research on pretrained vision--language transformers, following the work of~\citet{karpathy2015deep}, typically relies on readily available image captioning datasets such as COCO~\cite{lin2014microsoft} or Flickr30k \cite{young2014image} for evaluating their models~\cite{bleeker2022lessons,rao2022does} on retrieval tasks.
These datasets comprise complex images depicting multiple objects or actions in rich backgrounds.
They differ from object-centric datasets that focus on a single object of a specific category, such as birds in CUB Captions~\cite{welinder2010caltech}, clothing items in Fashion200~\cite{han2017automatic}, or medical imagery in CheXpert~\cite{zhang2022contrastive}. 
Despite the richness of visual domains in captioning datasets, their captions tend to be single-sentence descriptions of the objects or actions in the images or their attributes. 
In contrast, our dataset offers rich textual collections that provide external context for the images.

\paragraph{Wikipedia image--caption datasets.}
AToMiC is built on top of the WIT dataset~\cite{wit}.
The main distinctions between AToMiC and WIT are the evaluation settings and text collection scope. 
The WIT dataset provides publicly available image (URL)--text tuples, which can be used for a variety of tasks. 
For example, the Kaggle challenge hosted by the Wikimedia Foundation emphasizes image--caption matching,\footnote{\url{https://www.kaggle.com/c/wikipedia-image-caption}} while~\citet{kreiss-etal-2022-concadia} prioritize image caption generation. 
In contrast, our objective is to construct a retrieval dataset to support multimedia content creation.
Accordingly, our focus is on studying the image--document (section) associations and building AToMiC around \textit{ad hoc} retrieval tasks.
The relationship between AToMiC and WIT is summarized as follows:
\begin{itemize}[leftmargin=*]
    \item \textbf{Associations.} We focus on section-level image--text associations, following WIT's train/validation/test splits (see Section~\ref{sec:rel}).
    \item \textbf{Texts.} We crawl English Wikipedia articles and consider sections without associated images, creating a more authentic setting, as not every text segment should be associated with images. Additional details are available in Section~\ref{sec:texts}.
    \item \textbf{Images.} We reuse the images and their metadata from WIT but provide image pixel values that are stored in a unified format. Further details can be found in Section~\ref{sec:images}.
\end{itemize}

\section{Task Formulation}

In standard {\it ad hoc} retrieval, we assume the existence of a corpus $\mathcal{C}$ comprised of a collection of documents $\{d_1, d_2 \ldots d_n\}$.
In response to a user's information need represented as query $q$, the system's goal is to return a top-$k$ ranked list of documents that maximizes some metric of quality such as nDCG, MRR, or MAP.

AToMiC, as an image/text multi-modal retrieval test collection, has a slightly different setup.
It is composed of two separate collections:\ a collection of texts $\mathcal{C}_{T} = \{t_1, t_2 \ldots t_n\}$ ($t$ stands for text) and a collection of images $\mathcal{C}_{M} = \{m_1, m_2 \ldots m_n\}$ ($m$ stands for media).
Figure~\ref{fig:teaser} shows an example $t$ from $\mathcal{C}_{T}$ and an example $m$ from $\mathcal{C}_{M}$.
Each $m$ is comprised of the image itself (i.e., pixel values) {\it as well as} metadata (if available), which might include captions describing the image.
It's worth noting that Wikipedia captions are designed to connect images and their attached sections, as shown in the work of~\citet{kreiss-etal-2022-concadia} and the style guide.\footnote{\url{https://en.wikipedia.org/wiki/Wikipedia:Manual_of_Style/Captions}}
Thus, we expect the captions to be semantically related to their associated documents.
In the two tasks we define below, we leverage known image--text associations as labels.

\medskip
\noindent 
{\bf Image Suggestion Task (T2M).}
An information need (for convenience, a {\it query}), denoted $q$, is simply a text $t$ drawn from $\mathcal{C}_T$, i.e., $q \in \mathcal{C}_T$.
That is, an editor of Wikipedia examines {\it a specific section} of an article and wishes to add an appropriate image.
Given $q$, the system's task is to return a top-$k$ ranked list of images, drawn from $\mathcal{C}_{M}$ that maximizes some metric of quality.
Relevance in this context is operationalized as ``this image would be appropriate to attach to this section''.
So, for the query represented by the text $t$ about the popular activity in Half Dome, the image $m$ of the people climbing up the mountain in Figure~\ref{fig:teaser} would be relevant.
Note that in this case, the image is already associated with the section ``Ascents'' in the Wikipedia article on ``Half Dome''.

\medskip
\noindent 
{\bf Image Promotion Task (M2T).} 
This task is the inverse of the image suggestion task, where the query $q$ is an image from $\mathcal{C}_{M}$ and the collection to be searched is $\mathcal{C}_{T}$.
The goal is to return a top-$k$ ranked list of texts, where relevance is operationalized as ``this is a section of a Wikipedia article that would be an appropriate attachment point for this image''.
Although not the focus of our efforts, models built for this task could support authoring tools for crafting captions for images.
For example, the users can write a caption: ``Getting ready for Half Dome hike'' to make it relevant to the text section shown in Figure~\ref{fig:teaser}, or ``A mountaineer proceeds across the ridge'' to make the image relevant to the mountaineering article in Wikipedia.\footnote{\url{https://en.wikipedia.org/wiki/Mountaineering}}
The image promotion task requires retrieval systems to provide relevant contexts that can be used for downstream tasks such as descriptive image captioning~\cite{liu2021visual} and open-knowledge visual question answering~\cite{marino2019ok,schwenk2022okvqa}.

\section{Dataset Construction}
\label{sec:db}

In this section, we present AToMiC, a dataset designed for exploring image/text cross-modal retrieval in the context of multimedia content creation. 
The development of this dataset was driven by the following objectives:\ 
(1) ensure a diverse set of natural image--text associations; 
(2) provide a publicly available dataset with standardized evaluation procedures;
(3) incorporate a broad range of intra-modal diversity, as existing datasets primarily focus on short text captions and object-oriented images, whereas real-world texts encompass multiple fields, hierarchical relationships, and varied lengths, and image--caption datasets tend to consider only certain styles such as photographs; 
(4) provide a sufficient quantity of data for training and evaluating state-of-the-art neural networks.

\subsection{Relevance Judgments}
\label{sec:rel}

\paragraph{Building from WIT}
In order to create AToMiC, we leverage the Wikipedia-based Image Text (WIT) dataset~\cite{wit},\footnote{\url{https://github.com/google-research-datasets/wit}} which provides 37.6 million tuples consisting of image, text, and contextual data extracted from Wikipedia pages in 279 languages, last curated around August 30, 2020.
Contextual information includes section and page titles and descriptions, as well as captions, alt-text, and attribution descriptions for each image.
We utilize WIT tuples that correspond to section-level image--text associations, where an image is associated with a section of a page. 
To simplify evaluation, we restrict our attention to only the associations in English Wikipedia.

\paragraph{Defining sparse relevance labels.}
The relevance labels in our dataset, or qrels in TREC terminology, are based on image--section associations in English Wikipedia. 
As shown in the example snippet from the ``Half Dome'' Wikipedia page in Figure~\ref{fig:teaser}, a relevant pair is made of a section and its associated image.
By contrast, we define all other pairs as non-relevant, even if this leads to potential false negatives (cf. Section~\ref{sec:limit}).
We extract these pairs from WIT to form our AToMiC qrels and align the training, validation, and test splits used by the WIT dataset.\footnote{\url{https://github.com/google-research-datasets/wit/blob/main/DATA.md}}
These image--text pairs (qrels) are further filtered in three steps:
(1) We remove any associations where the image URL is no longer available on the Internet.
(2) We remap duplicate texts to the same id if their section and page titles and descriptions are the same. This step is necessary because the WIT image--text tuples are image-centric and the texts can be duplicated.
(3) We drop an image--text pair if either the image or text coexist in training, validation, and test sets to ensure there is no overlap. 
The reason is that we want to simulate the tasks of searching unknown images or texts.

\paragraph{Statistics.}
Table~\ref{tb:dataset} presents the statistics of our AToMiC dataset according to the qrels mentioned above.
Note that the number of ``queries'' in each split is indicated in the columns for training, validation, and test.
The ``other'' column indicates the texts and images not included in the three splits.
When conducting the image suggestion and image promotion tasks, the indexed and searched collections should correspond to the total number of texts $|C_{T}|$ or images $|C_{M}|$ in the three settings defined in Section~\ref{sec:setups}. 
Note that the many-to-many nature of the image--section associations is reflected in the M/T and T/M values displayed in Table~\ref{tb:dataset}.

\begin{table}[t]
\centering
\resizebox{.8\columnwidth}{!}{
\begin{tabular}{lrrrr}
\toprule
Split & Training & Validation & Test & Other \\
\midrule
\# texts  & 3,002,458 & 17,173 & 9,873 & 7,105,240 \\
\# images & 3,386,183 & 16,131 & 8,605 & 7,608,283 \\
\# qrels   & 4,401,903 & 17,801 & 9,873 & - \\
\# M/T    & 1.47 ($\pm$ 2.72) & 1.03 ($\pm$ 0.43) & 1.00 ($\pm$ 0.00) & - \\
\# T/M    & 1.30 ($\pm$ 0.82) & 1.10 ($\pm$ 0.43) & 1.15 ($\pm$ 0.53) & - \\
\bottomrule
\end{tabular}
}
\vspace{0.25cm}
\caption{
AToMiC dataset statistics. The number of texts/images is defined by the relevant labels. \# M/T (T/M) stands for the number of relevant images per text (texts per image).
}
\label{tb:dataset}
\vspace{-0.5cm}
\end{table}

\begin{table}[t]
\centering
\resizebox{.8\columnwidth}{!}{
\begin{tabular}{lccc}
\toprule
Field & Empty Rate & CLIP & BERT \\
\midrule
\multicolumn{4}{c}{AToMiC-Texts}\\
\midrule
Title (P)   &     0\% & 4.79 ($\pm$ 2.52)     & 4.68 ($\pm$ 2.23) \\
Title (S)   &  0.02\% & 2.53 ($\pm$ 3.17)     & 2.32 ($\pm$ 2.78) \\
Hierarchy   &     0\% & 2.30 ($\pm$ 8.05)     & 2.15 ($\pm$ 7.71) \\
Context (P) & 19.21\% & 146.15 ($\pm$ 159.54) & 137.81 ($\pm$ 150.99) \\
Context (S) &     0\% & 195.93 ($\pm$ 312.87) & 184.68 ($\pm$ 294.76) \\
\midrule
\multicolumn{4}{c}{AToMiC-Images}\\
\midrule
Reference  & 42.28\% & 17.46 ($\pm$ 23.65)  & 13.98 ($\pm$ 17.48) \\
Alt-text   & 74.79\% &  6.67 ($\pm$ 10.85)  &  5.90 ($\pm$ 8.90) \\
Attribute  &  2.57\% & 47.69 ($\pm$ 112.87) & 41.63 ($\pm$ 90.42)\\
\midrule
\multicolumn{4}{c}{COCO-Captions}\\
\midrule
Caption & 0\% & 11.54 ($\pm$ 2.70) & 11.81 ($\pm$ 2.81)\\
\bottomrule
\end{tabular}
}
\vspace{0.25cm}
\caption{
Statistics for the number of tokens.
We use the tokenizers of CLIP and BERT (uncased).
The number of tokens is counted without special tokens and empty instances.
}
\label{tb:text-stat}
\vspace{-0.5cm}
\end{table}

\subsection{Text Collection}\label{sec:texts}

We develop the AToMiC dataset with the aim of creating a realistic data distribution that better reflects the diversity of documents in the real world. 
To achieve this, we utilize the collection of English Wikipedia pages as our source of text data. 
Wikipedia aims to capture the whole of human knowledge, hence its pages are highly diverse with a wide range of topics, including biographies, politics, sports, education, geography, etc.

\paragraph{Text collection expansion.}
To account for the editorial consideration that not every text requires an associated image, we expand our text collection to include both texts with associated images and non-associated ones, which is not covered by the WIT dataset.
To achieve this, we conduct a projection step to expand the WIT dataset by mapping it to the latest Wikipedia text collection using the BM25 matching method employed in the work of~\citet{zhang2022making}. 
To this end, we parse Wikipedia using the latest XML dump and prepare its Lucene index for retrieval during the construction of the AToMiC dataset.\footnote{We parse the texts from the \texttt{enwiki-20221101} XML dump.} 
We then use the texts from WIT (around 3M instances) as queries to search the latest Wikipedia text collection, and we employ the Anserini toolkit to facilitate BM25 matching with $(k_{1}, b)=(0.9, 0.4)$. 
Finally, we select the top-1 ranked section and its related sections under the same Wikipedia page as candidates. 
Consequently, the projected text collection includes approximately 10M candidates in the training, validation, test, and other (without relevance labels) sets, as presented in Table~\ref{tb:dataset}.
This approach allows us to include a diverse set of texts, both with and without associated images, and expands the scope of our dataset to better reflect real-world scenarios.

\paragraph{Token statistics}
In this section, we examine the statistics of the key textual inputs from five fields: page (section and its hierarchy) titles denoted as title (P), title (S), and hierarchy, as well as page (section) context descriptions, denoted as context (P) and context (S).
As we consider mostly pretrained transformers, e.g., CLIP or BERT, the textual input needs to be tokenized, i.e. mapped into words/subwords that they accept as input. 
This is important, because we have a maximum token length for some models, for example, CLIP only accepts 77 tokens and will truncate anything larger.
The token statistics are reported in Table~\ref{tb:text-stat}. 
The table reveals that the number of tokens varies greatly across fields.
Descriptive fields such as context (S) and context (P) have the most tokens, with approximately 140 tokens for both CLIP and BERT, and a standard deviation of approximately 300, which could be a problem for CLIP. 
In contrast, abstractive fields such as title (P), title (S), and hierarchy contain only about 10 tokens with small standard deviations, which more closely resembles the statistics from COCO.

\begin{figure}
    \centering
    \includegraphics[width=.6\columnwidth]{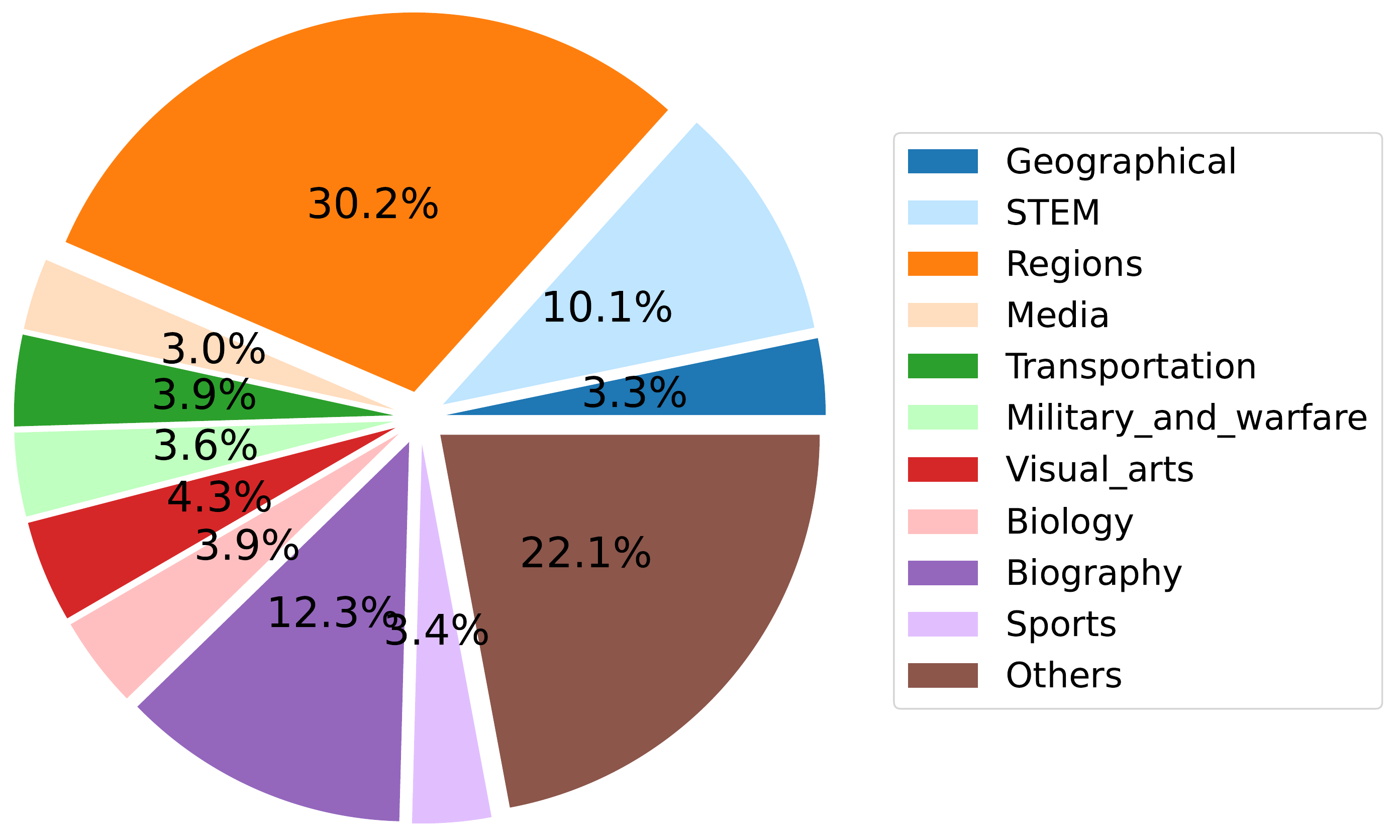}
    \caption{Wikipedia page category composition.}
    \label{fig:wiki_category}
\end{figure}

\paragraph{Wikipedia page categories.}
To showcase the diversity of our text collection, we use the predicted classes by the classifier from the Wikimedia Foundation.\footnote{Short link to the data: \url{https://tinyurl.com/pfjfa2e5}}
The category composition of English Wikipedia pages covered by our text collection is shown in Figure~\ref{fig:wiki_category}.
There are 34 predefined categories, and we combine those with less than 2.5\% as others in this figure.
As observed from the figure, the Wikipedia corpus is mostly about regions (30.2\%), Sports (12.3\%), and STEM (10.1\%) articles.
Other categories are distributed more equally: 3$\sim$4 \% per each.
There is also a long tail for other categories, where the proportion of the aggregated category is 22.1\% in total.

\subsection{Image Collection}\label{sec:images}

Our image collection is constructed by leveraging the high-quality and large-scale image collection available on Wikipedia, which includes a wide range of image styles beyond photographs, such as artworks, maps, radiographs, and more, making it a diverse and realistic resource compared to the existing image--caption datasets.

\paragraph{Image collection preparation.}
We retrieve all image URLs from the WIT dataset and used the \texttt{img2dataset} toolkit\footnote{\url{https://github.com/rom1504/img2dataset}} to crawl the images. 
Invalid URLs are filtered and the images were stored in WebP format with a quality parameter of 100.
Our image collection includes all images available in the WIT dataset, regardless of language, as dumping all images in the Wikipedia Commons database could be too aggressive for our task.

In addition to the images, we make use of other metadata from the WIT dataset, including contextualized captions in the reference, alt-text, and attribution fields.
Table~\ref{tb:text-stat} shows the token statistics for these fields.
When added to an article, this image metadata can make images more aligned with the context of the attached section.
Researchers could leverage this textual data to build reference systems using text-matching models.
However, it is worth noting that the empty rate for these captions is quite high, with approximately 42\% for the reference and 75\% for the alt-text fields.
The high empty rate of captions suggests that the generalization capability of matching systems relying on these features may be limited because they are not always available, especially in real-world scenarios where the images have not yet been linked to sections.

\begin{figure}[ht!]
\begin{subfigure}{.49\columnwidth}
\includegraphics[width=\columnwidth]{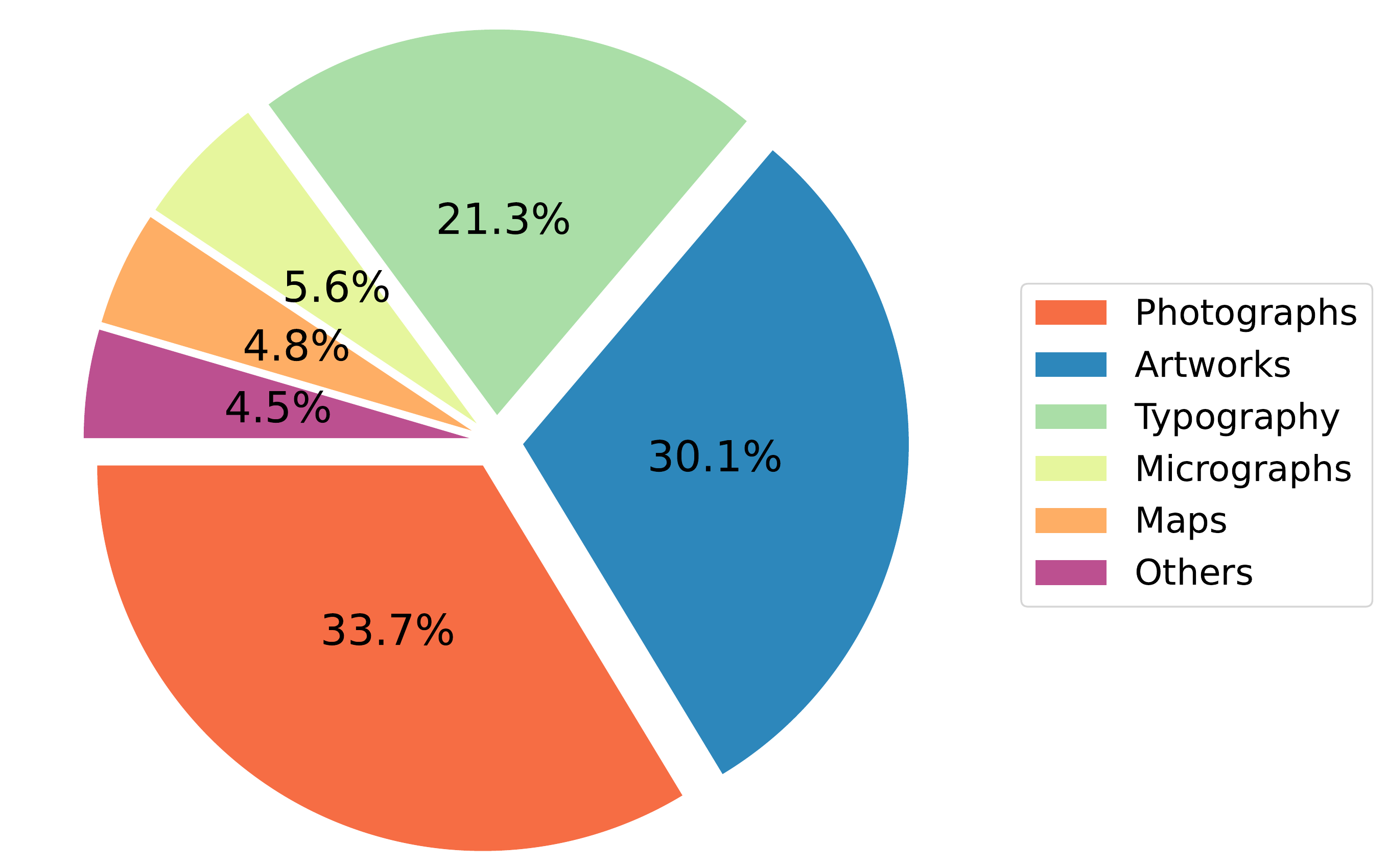}
\caption{AToMiC image styles}
\end{subfigure}
\begin{subfigure}{.49\columnwidth}
\includegraphics[width=\columnwidth]{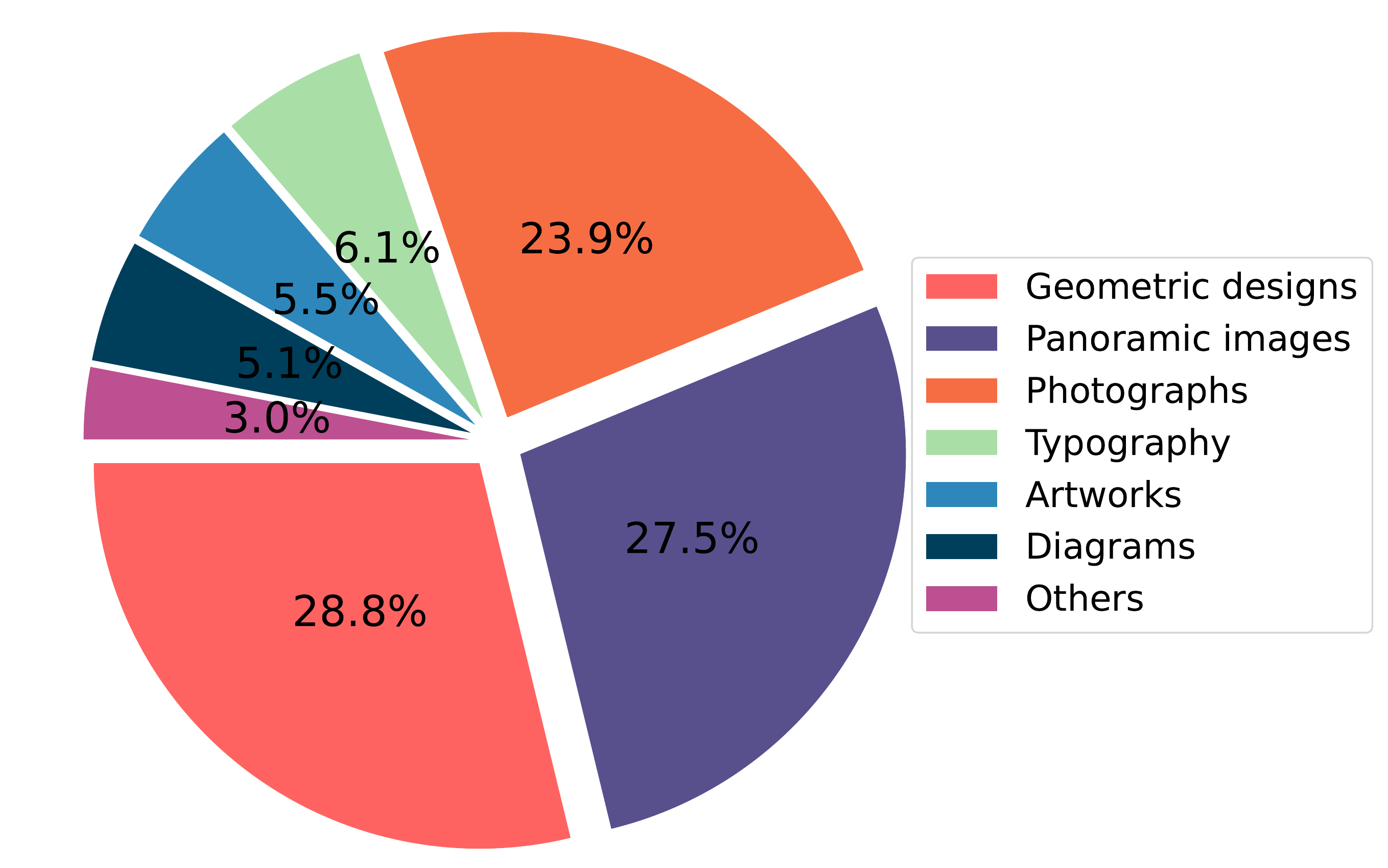}
\caption{COCO image styles}
\end{subfigure}
\caption{CLIP zero-shot image style analysis}
\label{fig:image_styles}
\vspace{-0.5cm}
\end{figure}

\paragraph{Image styles.}
We use CLIP to classify and analyze image styles in the AToMiC dataset.
Following~\citet{radford2021learning}, we encode the class embeddings for each of the 15 predefined image style text templates and conduct linear classification using the image embeddings encoded by CLIP.\footnote{\url{https://huggingface.co/openai/clip-vit-large-patch14}}
We construct the templates for image styles as the leading classes followed by their description with our heuristics.\footnote{Short link to the templates: \url{https://tinyurl.com/3fa8z6fn}}
We conduct zero-shot classification on our image collection as well as the images in the COCO dataset, and we combine the tail styles less than 1\% as others.
As shown in Figure~\ref{fig:image_styles}, AToMiC contains more artworks and maps compared to COCO. 
Note that the image styles might not be fully accurate because we are relying on CLIP's zero-shot image classification capability, which can be improved by providing more image-style labels or stronger models and left as future work.

\begin{figure}[ht!]
\begin{subfigure}{.49\columnwidth}
\includegraphics[width=\columnwidth]{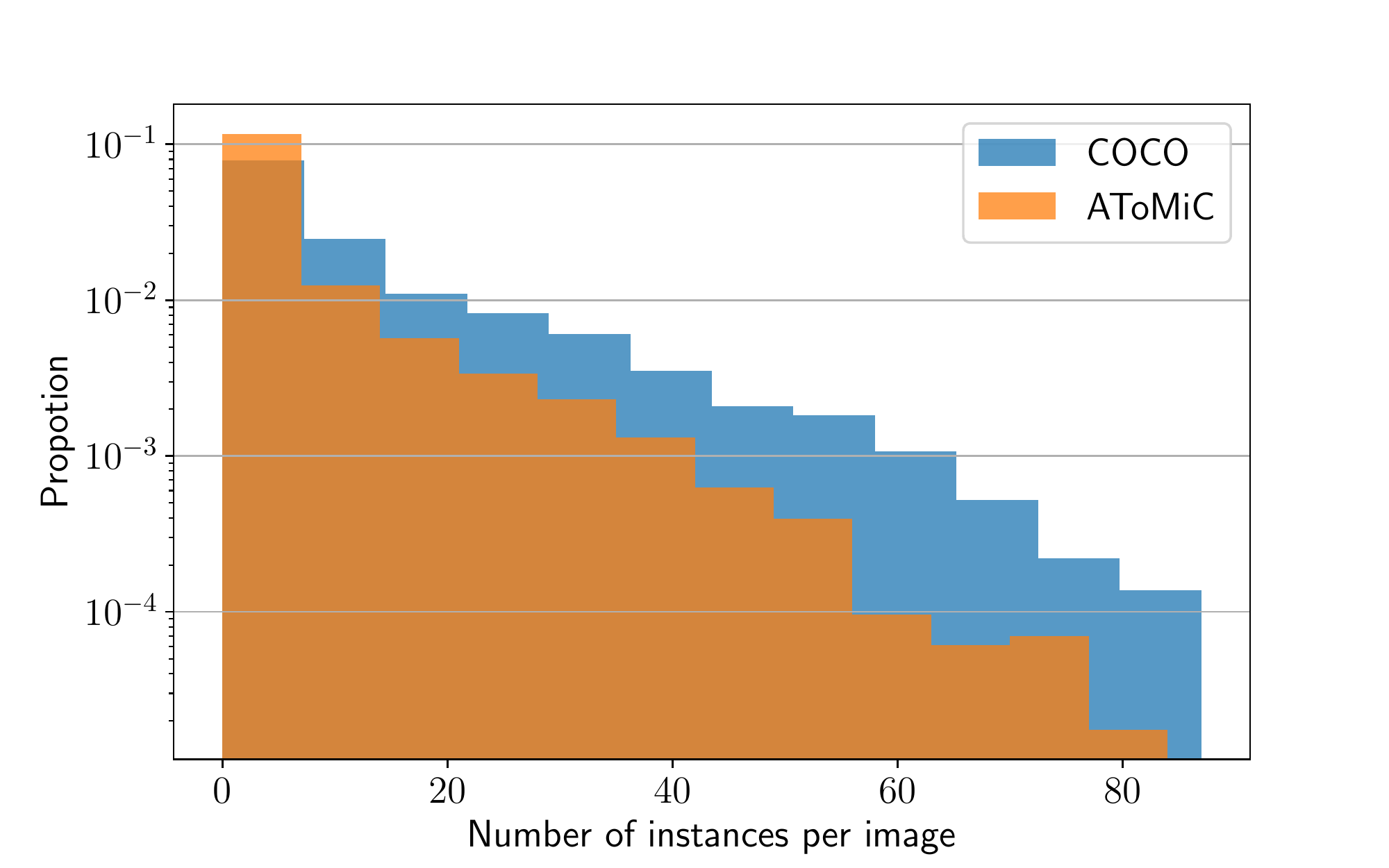}
\caption{Number of instances}
\end{subfigure}
\begin{subfigure}{.49\columnwidth}
\includegraphics[width=\columnwidth]{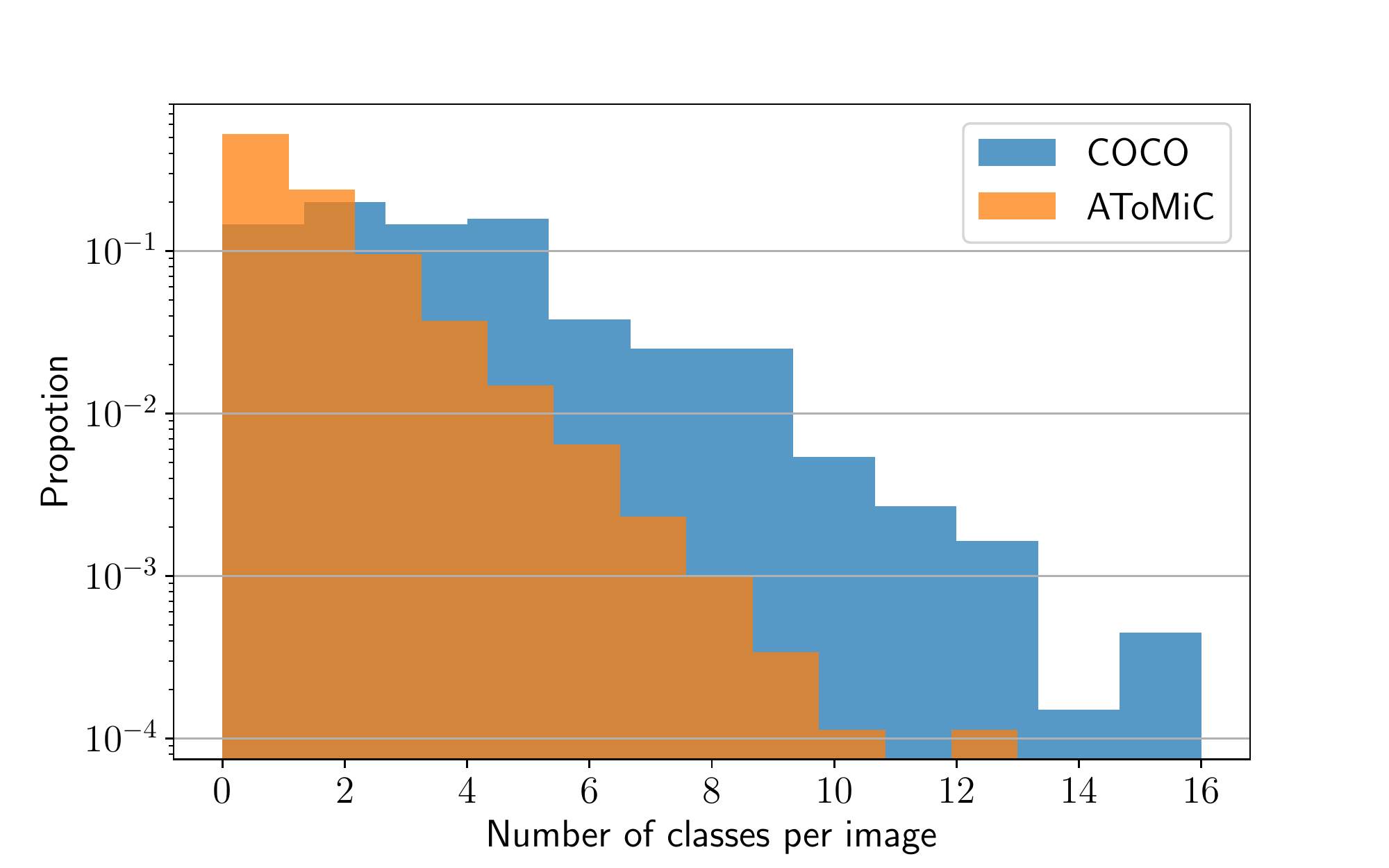}
\caption{Number of classes}
\end{subfigure}
\caption{DETR zero-shot objection detection analysis}
\label{fig:image_objects}
\vspace{-0.5cm}
\end{figure}

\paragraph{Object detection analysis.}
In addition to high-level information such as image classes, exploring the underlying objects in a scene is also important. 
Despite containing many non-photographic images, the AToMiC collection still includes a certain number of photograph-related images, as seen in Figure~\ref{fig:image_styles}. 
To conduct our preliminary analysis, we leverage the pretrained DETR~\cite{carion2020end} model to facilitate the object detection task for the AToMiC and COCO image collections.\footnote{\url{https://huggingface.co/facebook/detr-resnet-101}}
We count the number of instances and classes detected by DETR and present the results in histogram plots, as shown in Figure~\ref{fig:image_objects}.
The top-5 object classes in AToMiC are listed as follows: (1) person, (2) tie, (3) car, (4) chair, and (5) book.
Interestingly, observed from panel (a), the AToMiC image collection covers a wide range of the number of instances within an image, while the distribution of object classes of the AToMiC dataset is more centralized around a few object classes than that of COCO, as seen in panel (b). 
Note that DETR is trained on COCO, so there can be some bias in our evaluation.

\subsection{Evaluation Settings}
\label{sec:setups}

We present three settings based on the size of the corpus:
\begin{itemize}[leftmargin=*]
    \item \textbf{Small}: Candidates are limited to the splits defined in the dataset. 
    This is similar to the image--caption tasks~\cite{karpathy2015deep}, where the systems search for candidates in validation or test sets.
    The corpus size is around 20K, making it suitable for system development.
    \item \textbf{Base}: Candidates are expanded to include the training, validation, and test splits. 
    This enlarges the corpus by adding candidates from other splits, making it challenging for systems to distinguish the relevant items from the distractors. 
    The corpus size for this setting is around 3M, and it resembles the \textit{ad hoc} retrieval setting, but it only considers associated image--text pairs.
    \item \textbf{Large}: All candidates are searched, including the non-associated images and texts in the ``other'' set in Table~\ref{tb:dataset}. 
    The corpus size for both image and text retrieval tasks is around 10M. 
    This setup is more realistic for large-scale ad hoc retrieval, which makes it even more challenging for systems to determine whether an image should be attached to a section and vice versa.
\end{itemize}
We design these settings to reflect the difficulty of the retrieval tasks.
Current image--caption retrieval datasets have limited corpus sizes, which hinder their ability to replicate real-world retrieval scenarios. 
This limitation poses challenges for identifying differences between the latest models, as noted by~\citet{chun2021probabilistic,chun2022eccv_caption}. 
In response to this issue, researchers have attempted to combine arbitrary image--caption datasets to inject ``distractors''~\cite{geigle2022retrieve}.
However, the use of arbitrary distractors can result in undesired effects and make it difficult to compare, reproduce and interpret system effectiveness.
Our setup is designed explicitly to address this issue and to offer a more realistic setting for a large-scale cross-modal search engine.

Given the sparsity of the relevance labels, we use mean reciprocal rank (MRR@10) and recall (R@10, R@1K) as evaluation metrics.
We note that confusion can arise with the term Recall@k. 
Here we mean $\frac{1}{|Q|}{\sum_{q\in\{Q\}}\frac{\text{Retrieved positives@k}}{\text{Total positives}}}$, which differs from the definition used in the vision community: $\frac{1}{|Q|}{\sum_{q\in\{Q\}}\text{AtLeastOnePositive@k}}
$ \cite{hodosh2013framing,karpathy2015deep,faghri2018vse++}, for which we note as Success@k.
We use the \texttt{trec\_eval} and \texttt{ranx}~\cite{Bassani22ranx} toolkits to compute the metrics.

\begin{table*}
\resizebox{.95\textwidth}{!}{
\begin{threeparttable}[ht]
\begin{tabular}{lllllllll}
\toprule
\multicolumn{2}{c}{Base (Validation)} & \multicolumn{3}{c}{T2M} & \multicolumn{3}{c}{M2T} \\
Cond. & Model & MRR@10 & R@10 & R@1K & MRR@10 & R@10 & R@1K  \\
\midrule
\multicolumn{8}{c}{Oracle Method (contextualized captions as images)} \\
\midrule
(a) & BM25 & 0.285$^\text{bcdefghijk}$ & 42.16$^\text{bcdefghijk}$ & 70.76$^\text{bcdefghijk}$ & 0.431$^\text{bcdefghijk}$ & 57.90$^\text{bcdefghijk}$ & 80.73$^\text{bcdefghijk}$ \\
\midrule
\multicolumn{8}{c}{Noisy Image--Text Contrastive Pretraining} \\
\midrule
(b) & CLIP ViT-B-32 & 0.030$^\text{ijk}$ & 5.64$^\text{dijk}$ & 30.83$^\text{deijk}$ & 0.026$^\text{ijk}$ & 4.57$^\text{ijk}$ & 27.39$^\text{dijk}$ \\
(c) & CLIP ViT-L-14 & 0.069$^\text{bdeijk}$ & 12.25$^\text{dijk}$ & 47.34$^\text{deijk}$ & 0.044$^\text{bdijk}$ & 7.72$^\text{bdeijk}$ & 34.67$^\text{bdeijk}$ \\
(d) & OpenCLIP ViT-B-32$^{\dagger}$ & 0.030$^\text{ijk}$ & 4.57$^\text{ijk}$ & 22.78$^\text{ijk}$ & 0.031$^\text{bijk}$ & 5.26$^\text{bijk}$ & 24.25$^\text{ijk}$ \\
(e) & OpenCLIP ViT-B-32  & 0.038$^\text{bdijk}$ &  6.61$^\text{bdijk}$ & 29.63$^\text{dijk}$ & 0.041$^\text{bdijk}$ &  6.90$^\text{bdijk}$ & 31.82$^\text{bdijk}$ \\
(f) & OpenCLIP ViT-L-14 & 0.067$^\text{bdeijk}$ & 11.37$^\text{bdeijk}$ & 42.42$^\text{bdeijk}$ & 0.067$^\text{bcdeijk}$ & 11.46$^\text{bcdeijk}$ & 42.54$^\text{bcdeijk}$ \\
(g) & OpenCLIP ViT-H-14 & 0.076$^\text{bcdefijk}$ & 12.95$^\text{bdefijk}$ & 46.15$^\text{bdefijk}$ & 0.082$^\text{bcdefijk}$ & 13.61$^\text{bcdefijk}$ & 47.17$^\text{bcdefijk}$ \\
(h) & OpenCLIP ViT-G-14 & \textbf{0.087}$^\text{bcdefgijk}$ & \textbf{14.95}$^\text{bcdefgijk}$ & \textbf{50.06}$^\text{bcdefgijk}$ & \textbf{0.096}$^\text{bcdefgijk}$ & \textbf{15.84}$^\text{bcdefgijk}$ & \textbf{51.34}$^\text{bcdefgijk}$ \\
\midrule
\multicolumn{8}{c}{Image--Caption Multi-task Pretraining} \\
\midrule
(i) & BLIP ViT-B-32 & 0.008$^\text{k}$ & 1.42 & 12.46 & 0.010$^\text{k}$ & 1.84 & 14.95 \\
(j) & BLIP ViT-L-16 & 0.011$^\text{ik}$ & 2.20$^\text{ik}$ & 16.15$^\text{i}$ & 0.013$^\text{ik}$ & 2.44$^\text{ik}$ & 18.15$^\text{i}$ \\
(k) & FLAVA ViT-B-32 & 0.004 & 1.14 & 19.73$^\text{ij}$ & 0.007 & 1.62 & 21.24$^\text{ij}$ \\
\bottomrule
\end{tabular}
\begin{tablenotes}
    \item [$\dagger$] LAION-400m
\end{tablenotes}
\vspace{0.25cm}
\caption{
Zero-shot results using the AToMiC-Base setting for the image suggestion and image promotion tasks, which we denote as T2M and M2T, respectively.
Significance testing is performed using paired {\it t}-tests ($p < 0.05$) with Bonferroni correction $\alpha=55$.
}\vspace{-0.5cm}
\label{tb:baselines}
\end{threeparttable}
}
\end{table*}

\section{Retrieval Experiments}\label{sec:exp}

In this section, we present retrieval experiments that aim to address the following research questions:

\begin{enumerate}[label=RQ{\arabic*},leftmargin=*]
    \item How do state-of-the-art pretrained vision--language transformers perform in our retrieval tasks on the AToMiC test collection?
    \item Is their retrieval effectiveness robust to changes in index size?
    \item Can we effectively fine-tune the pretrained transformers using the provided training data?
    \item Do pretrained vision--language transformers capture useful information that is not captured by the textual features?
\end{enumerate}

\subsection{Baseline Results}
To answer the first research question, we conduct zero-shot retrieval experiments using publicly available implementations and model checkpoints.
To facilitate end-to-end retrieval when there are millions of candidates to be indexed and searched, we consider baselines that can encode images and texts independently.
Studies on more complex retrieval pipelines such as cascaded ranking are left for future work.

\paragraph{Baselines.}
We consider representative vision--language pretrained transformers that are publicly available and have the ability to encode image and text representations independently, i.e., bi-encoders.
In addition, we also consider BM25 using captions as our reference system.
Our baselines are as follows:

\begin{itemize}[leftmargin=*]
    \item Oracle Method (BM25): this is our reference system that utilizes contextualized captions to represent images instead of processing pixel values.
    We only consider captions in English and concatenate texts from the three fields:\ reference, alt-text, and attribution for images.
    For texts, we concatenate contents from the five fields in Table~\ref{tb:text-stat}.\footnote{[title (P), title (S), hierarchy, context (S),  context (P)]}

    \item Contrastive Pretraining: CLIP is a well-known vision--language pretrained transformer that can understand and generate natural language descriptions of images and classify images into a large number of categories.
    We consider both the implementation provided by the HuggingFace library and the open-sourced reproduction library, OpenCLIP.\footnote{\url{https://github.com/mlfoundations/open_clip}}
    The OpenCLIP library provides CLIP variants pretrained on the open-source LAION datasets~\cite{schuhmann2021laion,schuhmann2022laionb}; in contrast, the CLIP model released by OpenAI is pretrained on their proprietary dataset.
    With the consideration of further analysis in future work, we take OpenCLIP as our primary setup.
    We consider the OpenCLIP variants pretrained on the LAION-2B dataset that only contain image--text pairs in English without further specification.

    \item Multi-task pretraining: there has been significant progress in multimodal learning for image--text data, building upon the contrastive objective proposed by~\citet{radford2021learning}.
    The BLIP model~\cite{li2022blip} extends this by jointly learning from the contrastive, cross-attention ranking, and language modeling objectives.
    In contrast, the authors of the FLAVA model~\cite{singh2022flava} conduct a comprehensive study of additional pretraining objectives beyond contrastive learning, including masked image modeling, masked language modeling, and masked multimodal modeling.
    We utilize the open-source implementation and checkpoints available on HuggingFace for both FLAVA and BLIP models, which are fine-tuned on the COCO dataset.\footnote{\url{https://huggingface.co/facebook/flava-full}}\footnote{\url{https://huggingface.co/Salesforce/blip-itm-base-coco}}\footnote{\url{https://huggingface.co/Salesforce/blip-itm-large-coco}}
\end{itemize}

\noindent We perform end-to-end retrieval using Anserini (for BM25) and FAISS (for transformer-based models) by encoding image and text representations in a zero-shot manner. 
All transformer baselines use the same concatenated texts of the five fields for BM25 but pixel values for images. 
We analyze all baselines using the validation set without further specification.

\begin{figure*}[!ht]
\begin{center}

\begin{subfigure}[h]{1.5\columnwidth}
\centering
\adjustbox{max width=\textwidth}{%
\begin{tikzpicture}
\begin{customlegend}[
legend style={font=\tiny},
legend columns=5,
legend entries={
\textsc{BM25},
\textsc{ViT-B-32},
\textsc{ViT-L-14},
\textsc{ViT-H-14},
\textsc{ViT-G-14},
,
                        }]
         \addlegendimage{bmvTplot};
         \addlegendimage{vitbplot};
         \addlegendimage{vitlplot};
         \addlegendimage{vithplot};
         \addlegendimage{vitgplot};
        \end{customlegend}
\end{tikzpicture}
         }

\end{subfigure}

\begin{subfigure}[h]{0.99\columnwidth}
\adjustbox{max width=0.49\textwidth}{%
            \begin{tikzpicture}
       \begin{axis}[
           xlabel={Corpus Size},
           ylabel={MRR@10},
            xmode=log,
            log basis x={10},
            grid=major,
            minor tick num=0
            ]
           
         \addplot[bmvTplot] table {tikz_figs/corpus_size/t2i_mrr/bm25.txt};
         \addplot[vitbplot] table {tikz_figs/corpus_size/t2i_mrr/b.txt};
         \addplot[vitlplot] table {tikz_figs/corpus_size/t2i_mrr/l.txt};         
         \addplot[vithplot] table {tikz_figs/corpus_size/t2i_mrr/h.txt};
         \addplot[vitgplot] table {tikz_figs/corpus_size/t2i_mrr/g.txt};
         
       \end{axis}
    \end{tikzpicture}
         }
\adjustbox{max width=0.49\textwidth}{%
            \begin{tikzpicture}
       \begin{axis}[
           xlabel={Corpus Size},
           ylabel={Recall@1K},
            xmode=log, 
            log basis x={10},
                        grid=major,
            minor tick num=0
]
           
         \addplot[bmvTplot] table {tikz_figs/corpus_size/t2i_recall/bm25.txt};
         \addplot[vitbplot] table {tikz_figs/corpus_size/t2i_recall/b.txt};
         \addplot[vitlplot] table {tikz_figs/corpus_size/t2i_recall/l.txt};         
         \addplot[vithplot] table {tikz_figs/corpus_size/t2i_recall/h.txt};
         \addplot[vitgplot] table {tikz_figs/corpus_size/t2i_recall/g.txt};
         
       \end{axis}
    \end{tikzpicture}
         }
\caption{T2M retrieval}
\end{subfigure}
\begin{subfigure}[h]{.99\columnwidth}
\adjustbox{max width=0.49\textwidth}{%
            \begin{tikzpicture}
       \begin{axis}[
           xlabel={Corpus Size},
           ylabel={MRR@10},
            xmode=log,
            log basis x={10},
                        grid=major,
            minor tick num=0
]
           
         \addplot[bmvTplot] table {tikz_figs/corpus_size/i2t_mrr/bm25.txt};
         \addplot[vitbplot] table {tikz_figs/corpus_size/i2t_mrr/b.txt};
         \addplot[vitlplot] table {tikz_figs/corpus_size/i2t_mrr/l.txt};         
         \addplot[vithplot] table {tikz_figs/corpus_size/i2t_mrr/h.txt};
         \addplot[vitgplot] table {tikz_figs/corpus_size/i2t_mrr/g.txt};
         
       \end{axis}
    \end{tikzpicture}
         }
\adjustbox{max width=0.49\textwidth}{%
            \begin{tikzpicture}
       \begin{axis}[
           xlabel={Corpus Size},
           ylabel={Recall@1K},
            xmode=log,
            log basis x={10},             grid=major,
            minor tick num=0
]
           
         \addplot[bmvTplot] table {tikz_figs/corpus_size/i2t_recall/bm25.txt};
         \addplot[vitbplot] table {tikz_figs/corpus_size/i2t_recall/b.txt};
         \addplot[vitlplot] table {tikz_figs/corpus_size/i2t_recall/l.txt};         
         \addplot[vithplot] table {tikz_figs/corpus_size/i2t_recall/h.txt};
         \addplot[vitgplot] table {tikz_figs/corpus_size/i2t_recall/g.txt};
         
       \end{axis}
    \end{tikzpicture}
         }\caption{M2T retrieval}
\end{subfigure}
\caption{
Retrieval robustness against corpus size.
}
\label{fig:distractors}
\end{center}
\end{figure*}
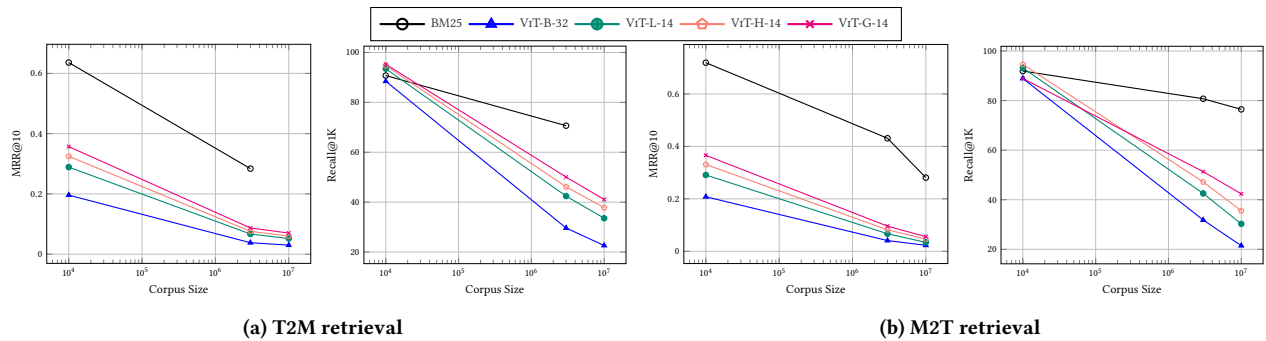

\paragraph{Observations.}
Zero-shot retrieval experiment results are shown in Table~\ref{tb:baselines}. 
BM25 (the oracle method) in row (a) demonstrates a high correlation between our sparse relevance labels and the contextualized captions, with an MRR@10 of around (0.284, 0.431) and R@1K of around (70.69, 80.74) for the image suggestion and promotion tasks, respectively.
While this method is more effective than transformer-based models, it is important to note that ``oracle'' captions for images are scarce in the real world.
Comparing rows (d) and (e), we observe marginal benefit of data scaling from LAION-400m to LAION-2B, when the vision encoder size is limited to ViT-B-32.
However, scaling up the vision encoder size to ViT-G-14 from rows (e) to (h) results in effectiveness improvements in R@1K, with scores of (50.06, 51.34).
While multi-task pretrained transformers typically outperform CLIP in the zero-shot retrieval setting using image--caption datasets, the results in rows (j)--(l) show that their effectiveness is lower than the basic CLIP settings in rows (b), (d), and (e) that use ViT-B-32 as their vision encoder, which is expected given that they are fine-tuned on the COCO dataset (which as we have seen is very different from AToMiC).

\begin{table}
\resizebox{\columnwidth}{!}{
\begin{threeparttable}[h]
\begin{tabular}{lcccrrrr}
\toprule
& \multicolumn{3}{c}{Base (Validation)} & \multicolumn{2}{c}{T2M} & \multicolumn{2}{c}{M2T} \\
Cond. & Titles$^{\dagger}$ & Context (S) & Context (P) & MRR@10 & R@1K & MRR@10 & R@1K \\
\midrule
(a) & \checkmark & \checkmark & \checkmark & 0.067 & 42.42 & 0.067 & 42.54\\
\midrule
(b) & \checkmark & & \checkmark & 0.053 & 32.76 & 0.052 & 34.52 \\
(c) & & \checkmark & & 0.054 & 38.66 & 0.057 & 39.11 \\
(d) & & & \checkmark & 0.017 & 15.70 & 0.025 & 20.86 \\
\bottomrule
\end{tabular}
\begin{tablenotes}
    \item [$\dagger$] Concatenated title (P), title (S), and hierarchy
\end{tablenotes}
\end{threeparttable}
}
\vspace{0.25cm}
\caption{
Ablation study of the usefulness of different fields using the OpenCLIP ViT-L-14 model
}\vspace{-0.8cm}
\label{tb:text-ablation}
\end{table}

\begin{table*}[t]
\centering
\resizebox{.8\textwidth}{!}{
\begin{tabular}{lccllllllll}
\toprule
\multicolumn{3}{c}{Base} & \multicolumn{4}{c}{Validation} & \multicolumn{4}{c}{Test} \\
\cmidrule(lr){4-7} \cmidrule(lr){8-11}
& & & \multicolumn{2}{c}{T2M} & \multicolumn{2}{c}{M2T} & \multicolumn{2}{c}{T2M} & \multicolumn{2}{c}{M2T}\\
Cond. & Method & Model & MRR@10 & R@1K & MRR@10 & R@1K & MRR@10 & R@1K & MRR@10 & R@1K \\
\midrule
(a) & \multirow{4}{*}{Zero shot} & BM25       & 0.285 & 70.76 & 0.431 & 80.73 & 0.359 & 78.12 & 0.447 & 83.29 \\
(b) &                 & ViT-B-32   & 0.038 & 29.61 & 0.041 & 31.80 & 0.038 & 29.15 & 0.044 & 31.91 \\
(c) &                 & ViT-L-14   & 0.067 & 42.42 & 0.067 & 42.54 & 0.066 & 42.40 & 0.067 & 42.85 \\
(d) &                 & ViT-G-14   & 0.087 & 50.06 & 0.096 & 51.34 & 0.089 & 51.19 & 0.095 & 52.73 \\
\midrule
\multicolumn{11}{c}{Fine-tuning; with $\alpha=6$ for (e,f,g,b) and (h,i,j,c)} \\
\midrule
(e) & Text         & \multirow{3}{*}{(b)} & 0.035 & 38.44$^\text{bf}$ & 0.044$^\text{bf}$ & 41.38$^\text{bf}$ & 0.037 & 39.66$^\text{bf}$ & 0.049$^\text{bf}$ & 44.06$^\text{bf}$\\
(f) & Vision       &  & 0.038$^\text{e}$ & 36.77$^\text{b}$ & 0.036 & 36.90$^\text{b}$ & 0.040 & 38.38$^\text{b}$ & 0.039 & 39.20$^\text{b}$\\
(g) & Text, Vision &  & 0.037 & 42.31$^\text{bef}$ & 0.043$^\text{b}$ & 43.80$^\text{bef}$ & 0.043$^\text{be}$ & 44.683$^\text{bef}$ & 0.0483$^\text{b}$ & 46.813$^\text{bef}$\\
\midrule
(h) & Text & \multirow{3}{*}{(c)} & 0.066 & 49.36$^\text{ci}$ & 0.074$^\text{cij}$ & 51.95$^\text{ci}$ & 0.063 & 50.13$^\text{c}$ & 0.075$^\text{cij}$ & 53.38$^\text{ci}$\\
(i) & Vision       & & 0.066 & 47.96$^\text{c}$ & 0.059 & 47.21$^\text{c}$ & 0.065 & 49.73$^\text{c}$ & 0.059 & 48.16$^\text{c}$ \\
(j) & Text, Vision & & 0.066 & 52.67$^\text{chi}$ & 0.068$^\text{i}$ & 53.63$^\text{chi}$ & 0.065 & 54.84$^\text{chi}$ & 0.071$^\text{i}$ & 56.05$^\text{chi}$\\
\midrule
\multicolumn{11}{c}{Fusion; with $\alpha=6$ for (a,d,k,l)}\\
\midrule
(k) &  RRF         &  \multirow{2}{*}{(a) + (d)} & 0.276$^\text{d}$ & \textbf{79.01}$^\text{ad}$ & 0.371$^\text{d}$ & \textbf{86.36}$^\text{ad}$ & 0.323$^\text{d}$ & \textbf{84.46}$^\text{ad}$ & 0.384$^\text{d}$ & \textbf{88.42}$^\text{ad}$ \\
(l) &  WSUM        &   & \textbf{0.329}$^\text{adk}$ & 78.95$^\text{ad}$ & \textbf{0.468}$^\text{adk}$ & 86.29$^\text{ad}$ & \textbf{0.402}$^\text{adk}$ & 84.45$^\text{ad}$ & \textbf{0.483}$^\text{adk}$ & 88.33$^\text{ad}$ \\
\bottomrule
\end{tabular}
}
\vspace{0.25cm}
\caption{
Fine-tuned models and fusion results.
Significance testing is performed using paired {\it t}-tests ($p < 0.05$) with Bonferroni correction $\alpha = 6$.
}
\label{tb:fusion}
\end{table*}

\paragraph{Ablation study for different text fields.}
Here, we investigate the usefulness of different text fields for our image--section retrieval task.
We conduct an ablation study using the OpenCLIP ViT-L-14 model and present our results in Table~\ref{tb:text-ablation}. 
We find that context (S) in row (c) provides the most useful information, with R@1K$=(38.66, 39.11)$ compared to R@1K$=(42.42, 42.54)$ in row (a) that takes all available information.
Other fields such as section and hierarchy titles provide limited information given their R@1K $< 10$.
We suggest that more sophisticated techniques can be developed to fuse text fields, which is left as future work.

\subsection{Retrieval Robustness vs.\ Corpus Size}

To address our second research question on system robustness and retrieval effectiveness against corpus size, we conducted a comparative analysis of OpenCLIP models and BM25 in the Small, Base, and Large settings, presented in Figure~\ref{fig:distractors}. 
Our results show that the OpenCLIP models perform exceptionally well when evaluated with the Small setting, achieving R@1K of around 90 for both tasks, comparable to BM25, which uses oracle features. 
However, while the off-the-shelf ViT-G-14 model can achieve MRR@10 of 0.35 without oracle features, its effectiveness deteriorates quickly as the corpus size increases. 
We observe a similar deterioration trend in the effectiveness of BM25, but its slope is smaller than that of the OpenCLIP models. 
It is worth noting that reporting BM25 scores for the T2M tasks in the Large setting could be misleading because the BM25 model did not consider non-English captions, i.e., no distractors were indexed.
We only employed BM25 as a reference for possible improvements in the future. 
The ultimate goal is to enable automatic systems to understand images without the aid of human-crafted oracle features.

\subsection{Fine-Tuning and Fusion}\label{sec:finetune}

Next, we address the final two research questions concerning the utility of (a) the training data and (b) the transformer-based models through fine-tuning and fusion experiments. 
As these experiments involve selecting hyperparameters, we present the test set results for reference. 
To facilitate comparison, we report the base runs in rows (a) to (d) of Table~\ref{tb:fusion}.

\paragraph{Fine-tuning settings.} 
We choose two OpenCLIP pretrained models:\ ViT-B-32 and ViT-L-14.
We use contrastive loss with in-batch negatives, following the work of~\citet{radford2021learning} but using the AToMiC training data.
The batch size is set to 1024 for all models, where we use the AdamW optimizer and train the models for 9075 steps with a learning rate of $10^{-5}$.
Following the work of~\citet{zhai2022lit}, we experimented with three settings when fine-tuning CLIP:\ (1) text encoder only (noted as Text); (2) vision encoder only (Vision); and (3) both of them (Text, Vision).

\paragraph{Fusion settings.}
Fusion techniques, e.g., early, late, or transmedia, are helpful for improving multimedia retrieval effectiveness~\cite{tois_crossmedia_2015}.
We use simple late fusion approaches by mixing the retrieved ranked list from BM25 and ViT-G-14:\ weighted sum (WSUM) with min-max normalization and reciprocal rank fusion (RRF).
The hyperparameters are optimized for MRR@10 on the validation set:\ the weights for WSUM are set to (BM25, ViT-G-14) $= (0.6, 0.4)$, and we set $k=30$ for RRF.

\paragraph{Observations.}
An overview of the table indicates that the evaluation results are consistent and highly correlated between the validation and test sets, confirming the observations made in the previous sections. 
Analysis of rows (e) to (j) reveals that the training data offers valuable information in terms of R@1K.
Comparison of row (e) to (g) with row (c) shows that the fine-tuned ViT-B-32 model performs similarly to the ViT-L-14 model in the zero-shot setting with respect to R@1K.
The same trend is observed for the fine-tuned ViT-L-14 models in rows (h) to (j). 
We mark two key observations regarding our fine-tuning approach: (1) it does not lead to improvements in MRR@10; (2) tuning both encoders consistently yields better results in terms of R@1K.

The results of our experiments show that while the BM25 baseline leverages oracle features, incorporating a transformer-based model via late fusion techniques can lead to significant improvements. 
This finding suggests that transformer-based models can still capture valuable information.
We observe enhancements in both MRR@10 and R@1K when using the WSUM approach, and although RRF falls short in MRR@10, it performs comparably to WSUM in R@1K.
We also compare the run coverage of BM25 with ViT-G-14, finding that BM25 covers only around 30\% of the candidates retrieved by ViT-G-14. 
This leads us to hypothesize that the pretrained OpenCLIP model is capable of capturing the missing information from images, as the information contained in images is dense and the contextualized captions may not be comprehensive enough to cover all aspects of an image.

\begin{figure*}[htb]
\begin{center}
\begin{subfigure}{0.3\columnwidth}
\includegraphics[width=\columnwidth]{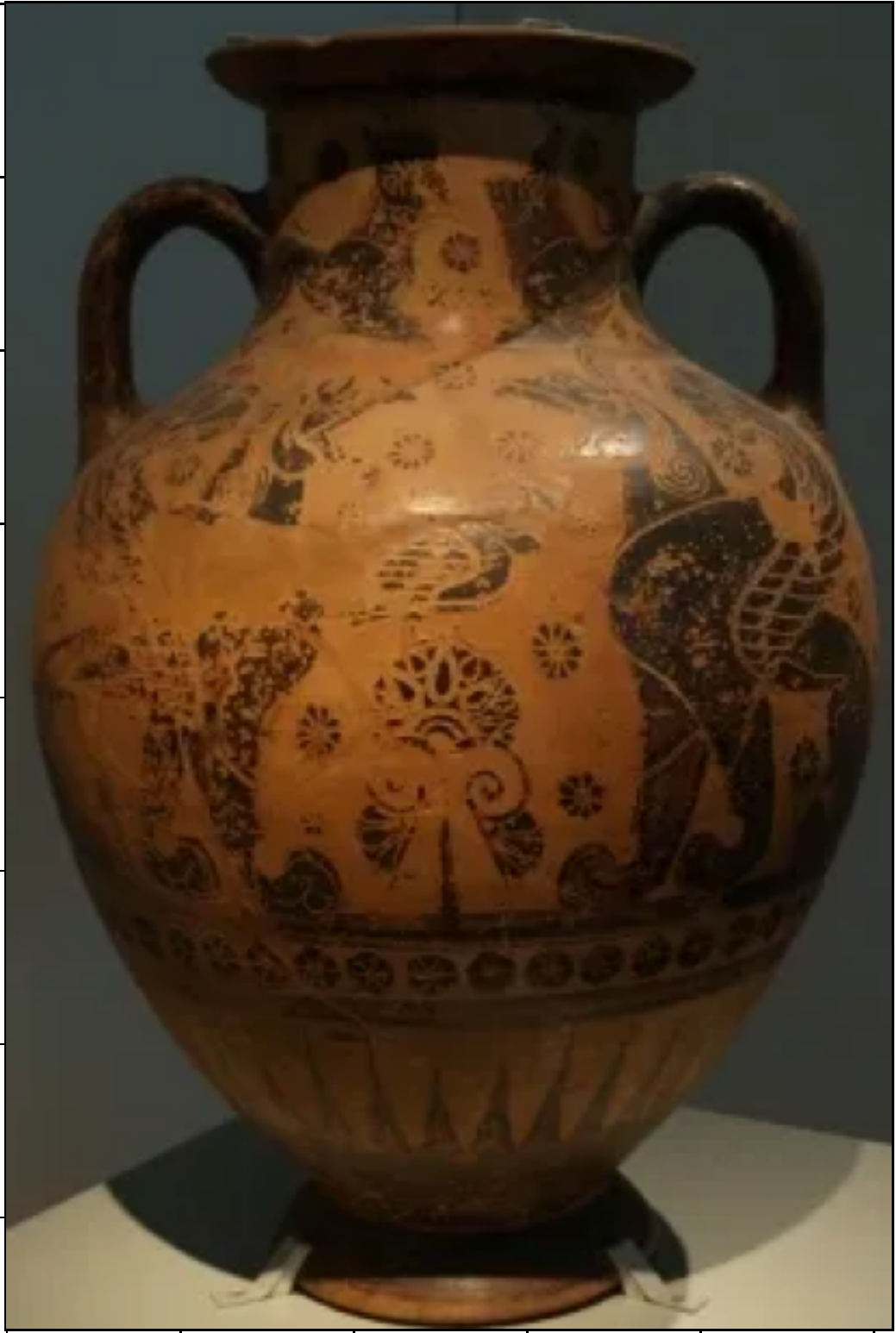}
\caption{Original attached image.}
\end{subfigure}
\quad
\begin{subfigure}{0.8\columnwidth}
\includegraphics[width=\columnwidth]{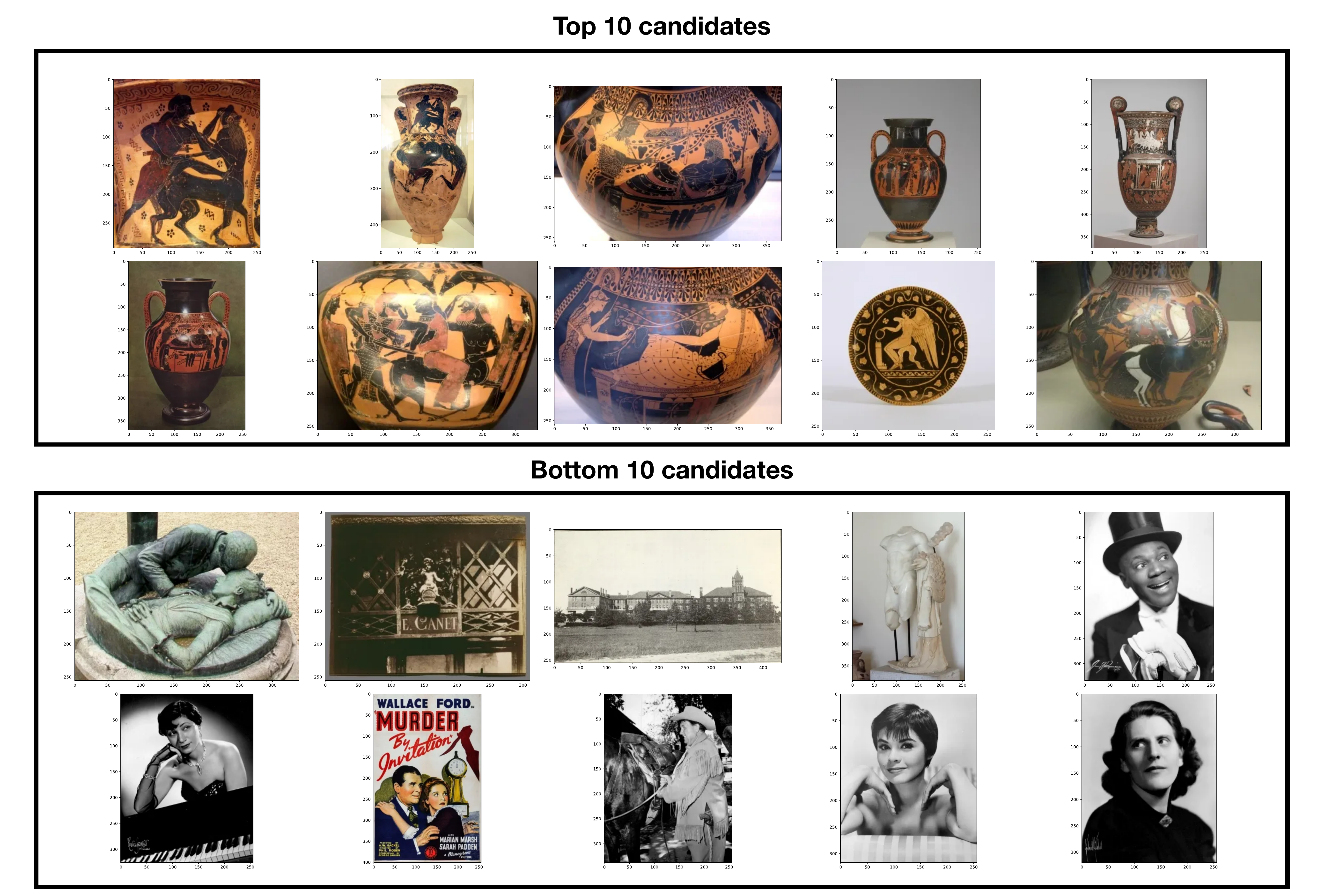}
\caption{Top-10 and Bottom-10 (out of 1K) candidates retrieved by BM25}
\end{subfigure}
\quad
\begin{subfigure}{0.8\columnwidth}
\includegraphics[width=\columnwidth]{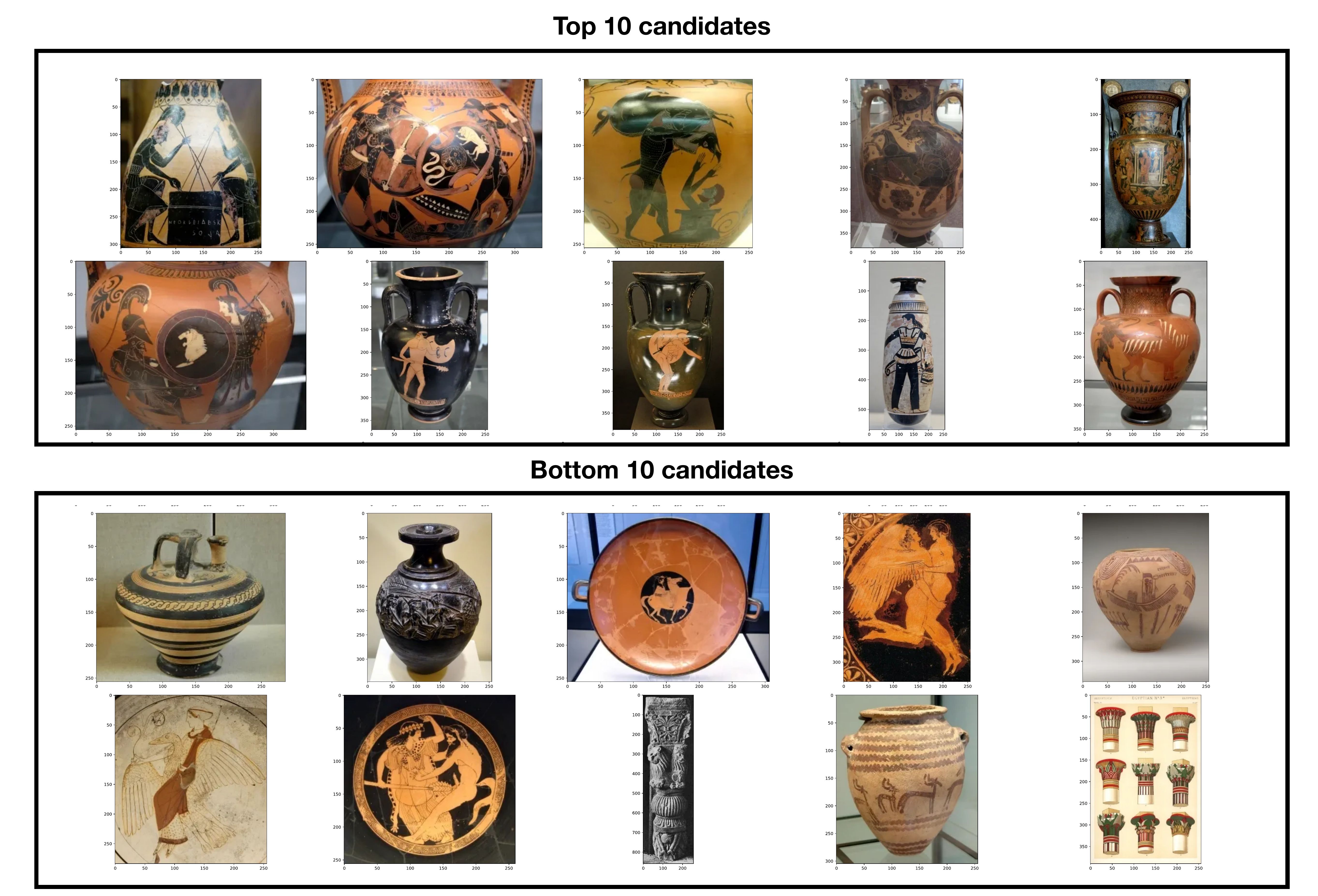}
\caption{Top-10 and Bottom-10 (out of 1K) candidates retrieved by OpenCLIP ViT-L-14}
\end{subfigure}
\end{center}

\vspace{0.1cm}

\begin{subfigure}{0.8\textwidth}
\begin{footnotesize}
\begin{verbatim}
'text_id': 'projected-06900598-003', 'page_title': 'Nessos Painter', 'section_title': 'Examples of work', 
'context_section_description': 
"In the name vase amphora depicting Nessos fighting Heracles, the painter utilizes iconography such as a depiction 
of with a mustache. This differs from artwork that typically shows Heracles with a beard and his usual 
attire of a lion skin cloak and lion mask. The names of both Nessos and Heracles are written above them, indicating 
..."
\end{verbatim}
\end{footnotesize}
\end{subfigure}
\caption{A case study comparing retrieved candidates.}
\label{fig:case}
\end{figure*}

\subsection{Case Study}

In this subsection, we present a qualitative analysis by investigating the retrieved candidates for a case study.
We show the retrieved images in Figure~\ref{fig:case}, where the text query is a section of ``Examples of Work'' from the page on Nessos Painter and is displayed below the figures. 
The original attached image, which is the ``ground truth'', is displayed in panel (a). 
We select the top 10 and bottom 10 (out of 1K) candidates in the list of retrieved results for both BM25 and OpenCLIP ViT-L-14, as illustrated in panels (b) and (c). 

Our findings suggest that the OpenCLIP model tends to search for similar artwork, whereas the BM25 model can already begin to focus on people at the bottom of the ranked list in this case.
This case exemplifies the fact that the image suggestion task requires greater background knowledge about the scene in the text, such as the leading clue in the section ``In the name vase amphora depicting Nessos fighting Heracles.''
OpenCLIP may be capable of identifying vases of related figures, but it struggles to make a precise decision given the context.

\section{Limitations and Future Work}\label{sec:limit}

When using the AToMiC dataset, it is important to be aware of its limitations.
Nevertheless, we believe that these limitations present an opportunity for the research community to engage in discussion and further exploration.
By doing so, we can collectively work towards improving the quality and utility of this dataset for future research endeavors.

\paragraph{Sparse judgments.}
AToMiC at present has a significant limitation due to the use of sparse proxy relevance labels.
The qrel collection is based on a highly biased sample of the universe of possible image--text relations.
Our experiments reveal that even with a substantial number of empty captions, BM25 with  captions delivers strong performance in terms of MRR@10 and R@1K.
However, it is unclear how this data bias relates to its relevance in the context of multimedia content creation.
Meanwhile, relying exclusively on captions could  introduce security issues such as ranking random images paired with the corresponding captions at the top, as the pixel-to-caption relations are disregarded.
Moreover, the sparse binary annotations in our dataset limit the use of metrics for further analysis such as MAP and nDCG.
To alleviate these issues, a more established evaluation is required, e.g., a TREC evaluation.

\paragraph{Popularity bias.}
The AToMiC dataset includes a diverse range of text and image data, which is drawn from the natural distribution of Wikipedia pages. 
However, this also introduces a potential bias, as the dataset is subject to popularity bias when evaluating systems using macro averaging over all the text and image ``queries.''
In our analysis of the text and image data, we have observed numerous long-tail cases that require automatic systems to improve their quality or utility. 
Marco-average metrics that consider all text and images may not adequately capture these long-tail cases. 
As such, future research should aim to develop and evaluate automatic systems that can address the specific needs of these long-tail cases, while also considering the potential impact of popularity bias on the effectiveness of these systems.

\paragraph{Static test collection.}
The AToMiC dataset captures a snapshot of Wikipedia that includes images, text, and their associations at the time when the dataset was curated. 
However, it is important to acknowledge that one key property of Wikipedia that we did not consider in our evaluations is that it is a time-varied collection.
As time passes, editors may choose to replace existing texts and images, which can impact the underlying data.
Furthermore, the image--text association that we captured in our dataset may also change if editors add or remove images from the pages.
As such, it is important to be aware of the dynamic nature of Wikipedia and the potential impact that changes to the underlying data can have on future research. 
Future work may benefit from considering the time-varying nature of Wikipedia when developing and evaluating automatic systems that rely on this dataset.

\paragraph{Lack of support for evaluating generative models.}
The recent advancements in text-to-image~\cite{pmlr-v139-ramesh21a,rombach2022high} and text-to-text generative models~\cite{gpt-3,t5} have shown great potential for enhancing creative digital content creation tasks.
However, the AToMiC dataset is limited in its support for evaluating such models. 
This is because all the images in the dataset need to be predefined and indexed for search purposes, which may not be conducive to evaluating the effectiveness of generative models that can sample images that do not exist in the real world. 
In certain scenarios, generated images or texts could potentially improve the quality and utility of the dataset for digital content creation tasks. 
Future research should consider the inclusion of such generative models in the evaluation to better assess their efficacy in real-world scenarios.

\section{Conclusion}\label{sec:conclusion}
In this paper, we have presented the AToMiC dataset, a realistic test collection designed to facilitate scalable, diverse, and reproducible multimedia retrieval research. 
We have introduced an approach for building the dataset that includes a realistic user model based on Wikipedia's text and image collections. 
The AToMiC dataset offers evaluation resources, baselines, and a community for studying the natural occurrences of texts and images.

Our experiments demonstrate the capabilities of the AToMiC dataset, contrasting it with the widely used COCO dataset. 
We formulate two retrieval tasks and conducted a preliminary evaluation using transformer-based models. 
The results show that the AToMiC dataset presents a more realistic depiction of the relationship between images and text in natural settings.

Overall, we provide AToMiC as a test collection for developing image--text retrieval systems aiming to assist multimedia content creation.
We believe that this dataset is helpful for image--text relation analysis and large-scale image--text retrieval benchmarks.
We encourage researchers to use the AToMiC dataset for future multimedia retrieval research and welcome further feedback and contribution. 
With the availability of this dataset, we anticipate that research communities will make significant strides in understanding image--text relationships and developing authoring tools for multimedia content creation.

\section*{Acknowledgements}

This research was supported in part by the Natural Sciences and Engineering Research Council (NSERC) of Canada.

\bibliographystyle{ACM-Reference-Format}
\bibliography{ref}

\clearpage

\appendix

\pgfplotsset{every tick label/.append style={font=\small}}

\begin{figure}[!ht]
\begin{center}

\begin{subfigure}[h]{.9\columnwidth}
\centering
\adjustbox{max width=0.7\textwidth}{%
\begin{tikzpicture}
\begin{customlegend}[
legend style={font=\tiny},
legend columns=5,
legend entries={
\textsc{OpenCLIP},
\textsc{OpenCLIP (ft)},
\textsc{CLIP},
\textsc{BLIP},
\textsc{FLAVA},
,
                        }]
         \addlegendimage{openclipplot};
         \addlegendimage{openclipFplot};
         \addlegendimage{clipplot};
         \addlegendimage{blipplot};
         \addlegendimage{flavaplot};
        \end{customlegend}
\end{tikzpicture}
         }

\end{subfigure}

\begin{subfigure}[h]{.49\columnwidth}
\adjustbox{max width=\textwidth}{%
            \begin{tikzpicture}
       \begin{axis}[
           xlabel={\small{COCO (S@10)}},
           ylabel={\small{AToMiC (R@10)}},
           ymin=10, ymax=80,
           xmin=60,xmax=100,
           ]
         \addplot[openclipplot] table {tikz_figs/atomic_coco/t2i/openclip.txt};
         \addplot[openclipFplot] table {tikz_figs/atomic_coco/t2i/openclipf.txt};
         \addplot[clipplot] table {tikz_figs/atomic_coco/t2i/clip.txt};         
         \addplot[blipplot] table {tikz_figs/atomic_coco/t2i/blip.txt};
         \addplot[flavaplot] table {tikz_figs/atomic_coco/t2i/flava.txt};
         
       \end{axis}
    \end{tikzpicture}
         }
\caption{T2M correlation}
\end{subfigure}
\begin{subfigure}[h]{.49\columnwidth}
\adjustbox{max width=\textwidth}{%
            \begin{tikzpicture}
       \begin{axis}[
           xlabel={\small{COCO (S@10)}},
           ylabel={\small{AToMiC (R@10)}},
           ymin=10, ymax=80,
           xmin=80,xmax=100,
           ]
         \addplot[openclipplot] table {tikz_figs/atomic_coco/i2t/openclip.txt};
         \addplot[openclipFplot] table {tikz_figs/atomic_coco/i2t/openclipf.txt};
         \addplot[clipplot] table {tikz_figs/atomic_coco/i2t/clip.txt};         
         \addplot[blipplot] table {tikz_figs/atomic_coco/i2t/blip.txt};
         \addplot[flavaplot] table {tikz_figs/atomic_coco/i2t/flava.txt};
         
       \end{axis}
    \end{tikzpicture}
         }
\caption{M2T correlation}
\end{subfigure}
\caption{
Effectiveness comparison. We report R@10 for AToMiC and S@10 for COCO using transformer-based baselines and the OpenCLIP models fine-tuned on AToMiC.
}
\label{fig:correlation}
\end{center}
\vspace{-0.2cm}
\end{figure}
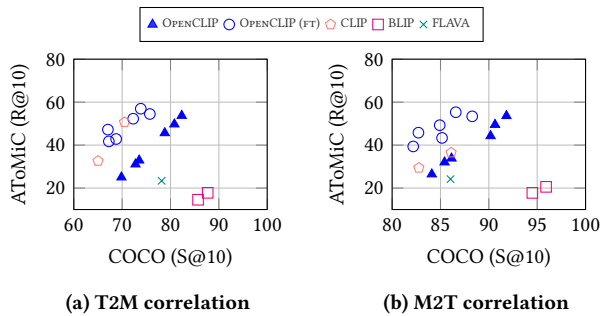

\section{AToMiC vs. COCO}
In this subsection, we extend our analysis of model retrieval effectiveness on the AToMiC and COCO datasets. 
For evaluation of the COCO dataset, we employ the Success@10 (S@10) metric using the Small setting on the test set in accordance with the common setting used in previous research. 
Similarly, for AToMiC, we report R@10 on the AToMiC validation set using the Small setting. 
Since the COCO dataset lacks oracle features owing to its image--caption retrieval setting, we compare only transformer-based models. 
We present the T2M and M2T results as scatter plots in Figure~\ref{fig:correlation}. 
Our findings suggest that the effectiveness of FLAVA and BLIP models is more skewed toward COCO but worse on AToMiC. 
Conversely, the CLIP and OpenCLIP models exhibit more neutral effectiveness for the two datasets. 
We hypothesize that the reason behind this is that FLAVA and BLIP models are trained on the aggregated image--caption datasets, which is very different from our task setting. 
The skewed effectiveness is particularly apparent in the BLIP models, where the authors proposed to further train their model on the bootstrapped image--caption pairs generated by itself. 
Lastly, our fine-tuning recipes still have room for improvement. 
As seen in the figure, our fine-tuned OpenCLIP models perform worse on COCO.
This could be attributed to the catastrophic forgetting phenomenon in our continuous learning setup.

\section{Image Analysis}
The images are stored as bytes of \texttt{Pillow} object in the \texttt{image} column.
To facilitate the image analysis in Section~\ref{sec:images}, users can leverage the \texttt{Transformers} library.\footnote{Documentation avaliable at \url{https://huggingface.co/docs/transformers/index}}
We also provide examples of using pretrained models such as CLIP and DETR for image-style classification and object detection tasks, respectively.

\subsection{Image-style Classification}
For the image-style classification task, we utilized the approach proposed by \citet{radford2021learning} and employed CLIP as a zero-shot linear classifier. 
We curated a set of 15 image styles in the format of \texttt{<image style>: <description>} and added two prompts, \texttt{an image of} and \texttt{a rendering of}, before each image-style class. 
We encoded the weights for each class using the CLIP text encoder, where each weight vector is the average of l2 normalized vectors of the same image-style class with two prompts. 
For a detailed example of the image-style classification task, refer to the notebook.\footnote{\url{https://github.com/TREC-AToMiC/AToMiC/blob/f9c6ac86dc0baac5e8c84f3fdf75a37e0022af1a/notebooks/02_Image_Styles.ipynb}}

\subsection{Object Detection}
To perform the object detection task, we employed the pre-trained DETR model. 
We used this model to process the images and obtain the number of instances and classes detected. 
A reference notebook detailing this process can be found in our repository in our repo.\footnote{\url{https://github.com/TREC-AToMiC/AToMiC/blob/f9c6ac86dc0baac5e8c84f3fdf75a37e0022af1a/notebooks/03_Object_Detection.ipynb}} 
As a preprocessing step to design a multi-stage image--text retrieval pipeline, users may utilize the bounding boxes to extract regional features and then build the matching model on top of them~\cite{geigle2022retrieve}.

\end{document}